\documentclass[adraft,copyright,creativecommons]{eptcs}

\usepackage{physics}
\usepackage{amssymb}
\usepackage{iftex}
\usepackage{amsthm}
\usepackage{stmaryrd}
\usepackage{hyperref}
\usepackage{ulem}
\usepackage{tikz}
\usepackage{tikzit}
\usepackage{float}
\usepackage{wrapfig}
\definecolor{zx_red}{RGB}{232, 165, 165}
\definecolor{zx_green}{RGB}{216, 248, 216}
\definecolor{zx_grey}{RGB}{211,211,211}
\definecolor{dzx_grey}{RGB}{150,150,150}
\usetikzlibrary{arrows.meta}

\tikzstyle{hadamard}=[fill=yellow, draw=black, shape=rectangle, tikzit draw=black, minimum size=5pt, inner sep=1.5pt ]
\tikzstyle{noeud}=[rounded rectangle, font={\scriptsize},inner sep=2pt, minimum width=0.7em, minimum height=0.7em, fill opacity=.7]
\tikzstyle{vert}=[style=noeud, fill={zx_green}, draw=black, tikzit fill=green]
\tikzstyle{dvert}=[style=noeud, fill={zx_green}, draw={dzx_grey}, tikzit fill=green]
\tikzstyle{drouge}=[style=noeud, fill={zx_red}, draw={dzx_grey}, tikzit fill=red]
\tikzstyle{rouge}=[style=noeud, fill={zx_red}, draw=black, tikzit fill=red]
\tikzstyle{svert}=[style=dvert, thick, tikzit fill=green]
\tikzstyle{srouge}=[style=drouge, thick, tikzit fill=red]
\tikzstyle{scalar}=[style=noeud, fill={zx_grey}, tikzit fill=gray]
\tikzstyle{mat}=[shape=signal, fill={zx_grey}, draw={dzx_grey}, minimum height=6pt, inner sep=1pt, font={\scriptsize\boldmath}, tikzit fill=gray, anchor=center, outer sep=-.1cm, signal pointer angle=120, scale=1]
\tikzstyle{lmat}=[style=mat, signal to=west, signal from=east]
\tikzstyle{rmat}=[style=mat, signal to=east, signal from=west]
\tikzstyle{arrow}=[->]
\tikzstyle{tarrow}=[->>]
\tikzstyle{grey}=[-, draw={zx_grey}, draw opacity= 0.8, tikzit draw=gray]
\tikzstyle{dgrey}=[-, draw={dzx_grey}, tikzit draw=gray]
\tikzstyle{gros}=[-,ultra thick, draw={zx_grey}, tikzit draw=red]

\tikzstyle{harrow}=[{Hooks[right]}->]
\tikzstyle{tharrow}=[{Hooks[right]}->>]

\newcommand{\interp}[1]{\left\llbracket#1\right\rrbracket}

\newtheorem{definition}{Definition}
\newtheorem{proposition}{Proposition}
\newtheorem{theorem}{Theorem}

\ifpdf
\usepackage{underscore}         
\usepackage[T1]{fontenc}        
\else
\usepackage{breakurl}           
\fi

\title{The decohered ZX-calculus}
\author{Titouan Carette \qquad Daniela Cojocaru
	\institute{LIX, CNRS, École polytechnique, Institut Polytechnique de Paris,  91120 Palaiseau, France}
	\email{titouan.carette@polytechnique.edu}
	\and
	Renaud Vilmart
	\institute{Université Paris-Saclay, CNRS, ENS Paris-Saclay, Inria, LMF, 91190, Gif-sur-Yvette, France}
	\email{renaud.vilmart@inria.fr}
}
\def\titlerunning{The decohered ZX-calculus}
\def\authorrunning{T.Carette, D. Cojocaru \& R. Vilmart}

\newcommand{\bvdots}{ \tikz[baseline] \node foreach \x in {-5,0,5} at (0,\x pt) [inner sep=0]{.};}

\newcommand{\sbvdots}{ \tikz[baseline,scale=0.8] \node foreach \x in {-6,0,6} at (0,\x pt) [inner sep=0]{.};}

\newcommand{\dec}[1]{#1^\circ}
\newcommand{\disc}{\mathbf{d}}

\begin{document}
	\maketitle
	
	\begin{abstract}
		The discard ZX-calculus is known to be complete and universal for mixed-state quantum mechanics, allowing for both quantum and classical processes. However, if the quantum aspects of ZX-calculus have been explored in depth, little work has been done on the classical side. In this paper, we investigate a fragment of discard ZX-calculus obtained by decohering the usual generators of ZX-calculus. We show that this calculus is universal and complete for affinely supported probability distributions over $\mathbb{F}_{2}^{n}$. To do so, we exhibit a normal form, mixing ideas from the graphical linear algebra program and diagrammatic Fourier transforms. Our results both clarify how to handle hybrid classical-quantum processes in the discard ZX-calculus and pave the way to the picturing of more general random variables and probabilistic processes.
	\end{abstract}
	
	\section{Introduction}\label{sec:intro}
	
	The ZX-calculus has been introduced in \cite{coecke2008interacting} as a convenient alternative to the matrix formalism of quantum mechanics, taking the radical stance of reformulating all concepts as string diagrams. String diagrams are formal combinatorial objects, originating from category theory, that are at the same time intuitive representations and reasoning tools through local rewrite rules \cite{piedeleu2023introduction}. Since then, ZX-calculus has been developed in various directions finding use cases in quantum compilation \cite{duncan2020graph} or simulation \cite{kissinger2022simulating}. Variations of ZX-calculus have been developed for qudits \cite{wang2021qufinite} or mixed states \cite{carette2021completeness}. In this last case, the so-called discard ZX-calculus is universal and complete for completely positive maps between Hilbert spaces of dimension $2^n $. Completely positive maps are the correct mathematical setting to represent probabilistic mixtures of pure quantum processes. So far, no particular attention has been given to the fragment of discard ZX-calculus corresponding to classical probabilistic processes. The goal of the present work is to fill this gap by investigating how the structures and generators of the discard ZX-calculus behave in the decohered setting, meaning when systems have lost all quantum features. One could be intrigued by the idea of using ZX-calculus in a non-quantum situation, however, we hope to show that the fundamental approach of ZX-calculus, relying on two interacting spider generators, is still relevant in the probabilistic setting. This leads to compact representations of probability distributions together with non-trivial rewriting rules paving the way towards diagrammatical probabilistic reasoning.
	
	The first contribution of this paper is the introduction of a notion of decohered maps encoding classical computations in quantum processes. We show that those decohered maps are in bijection with matrices with positive real coefficients. Our second contribution is the design of a decohered version of ZX-calculus obtained by decohering the generators of the discard ZX-calculus. We provide sound rewriting rules for this calculus and show that it is universal and complete for matrices with positive real coefficients having affine support. This universality and completeness results are the main achievement of this paper and rely on a new normal form inspired by previous works on diagrammatic Fourier transform \cite{kuijpers2019graphical} and scalable notations \cite{carette2019szx}.
	
	The paper is organized as follows. Section \ref{sec:mixed} recalls the necessary notions of mixed-state quantum mechanics, especially density matrices and completely positive maps. It also introduces the decohered maps our calculus aims to depict. Section \ref{sec:disczx} presents the generators and equations of the discard ZX-calculus of \cite{carette2021completeness}. Then, Section \ref{sec:deczx} properly introduces the decohered ZX-calculus, discussing in detail its rules and generators. Finally, in Section \ref{sec:unicomp}, we show our main results about the structure of decohered ZX-calculus, introducing our normal form and proving universality and completeness for matrices with positive real coefficients having affine support. All proofs can be found in Appendix \ref{append}.
	
	\section{Mixed States Quantum Mechanics}\label{sec:mixed}
	
	In this section, we recall the basics of density matrices and completely positive maps \cite{nielsen2010quantum}. We also introduce the notion of decohered maps.
	
	\subsection{Completely positive maps}
	
	In this paper, we are principally interested in qubit processing, we will write $n\to m$ to denote a quantum process from $n$ to $m$ qubits. Registers of $n$ qubits are represented by density matrices of size $2^n \times 2^n $.
	
	\begin{definition}[Density matrix]
		A \textbf{density matrix} for $n$ qubits is a matrix $\rho\in \mathcal{M}_{2^n \times 2^n}(\mathbb{C})$ that is Hermitian, positive, and has trace one, i.e. $\rho^\dagger = \rho $, there is a matrix $M$ such that $\rho= MM^\dagger $, and $\tr(\rho)=1$.
	\end{definition}
	
	Given a register of qubits represented by a density matrix $\rho$, the probability of observing the binary word $x$ is given by $\mathbb{P}(x)=\bra{x}\rho\ket{x}$. The positivity condition ensures that this number is a positive real number and the trace condition ensures that we have a well normalised probability distribution. Quantum processes are represented by CPTP maps.
	
	\begin{definition}[CPTP map]\label{def:CP}
		A \textbf{Completely Positive} (CP) map is a linear map $f: \mathcal{M}_{2^n \times 2^n}(\mathbb{C})\to \mathcal{M}_{2^m\times 2^m}(\mathbb{C})$ which is positive, meaning that it maps positive matrices to positive matrices, and such that for any $k$, $f\otimes I_{k}$ is positive. A CP map is said \textbf{Trace Preserving} (CPTP) if it preserves the trace.
	\end{definition}
	
	The requirements of CPTP maps are exactly what is needed to map density matrices to density matrices and be stable by tensor product. Density matrices can be identified with the CPTP maps $\mathbb{C} \to \mathcal{M}_{2^n\times 2^n}(\mathbb{C})$. An important class of CP maps are the pure maps.
	
	\begin{definition}[Pure map]
		A \textbf{pure map} $\underline{A}:n\to m$ is determined by a matrix $A\in \mathcal{M}_{2^m \times 2^n}(\mathbb{C})$ and defined as $\underline{A}:\rho \mapsto A\rho A^\dagger $. A pure map is always CP and CPTP if $A^\dagger A = I $, exactly when $A$ is an isometry.
	\end{definition}
	
	Pure maps are exactly the image of an embedding of complex matrices into CP maps. Indeed, $\underline{I_{n }}=id_n $, $\underline{A}\circ \underline{B}=\underline{AB}$ and $\underline{A\otimes B}=\underline{A} \otimes \underline{B}$. This embedding is not faithful but not far, as $\underline{A}=\underline{B}$ if and only if $A=wB$ for $w\in \mathbb{C}$ and $|w|=1$. There are CP maps that are not pure, a typical example is the discard map.
	
	\begin{definition}[Discard map]
		The \textbf{discard map} $\disc_n : n\to 0$ is defined as $\disc_n :\rho \mapsto \tr(\rho)$.
	\end{definition}
	
	The discard map is CPTP. We talk of a partial trace when we apply the discard map tensored with the identity. The general structure of CP maps is well understood.
	
	\begin{proposition}\label{prop:cpmstruct}
		Any CP map $f:n\to m$ is of the form $(\disc_k \otimes id_m )\circ \underline{A}$. The matrix $A\in \mathcal{M}_{2^{k+m}\times 2^n}$ is called a \textbf{purification} of $f$. Purifications are unique up to isometries,
		$(\disc_k \otimes id_m )\circ \underline{A}=(\disc_l \otimes id_m )\circ \underline{B}$, with $k\leq l$ if and only if there is a $V\in \mathcal{M}_{2^{l}\times 2^k}$ such that $V^\dagger V = I_{2^k}$ and $(V\otimes id_m )\circ A= B$.
	\end{proposition}
	
	Thus, any CP map is a pure map followed by a partial trace. CPTP are exactly the CP maps whose purifications are isometries.
	
	\subsection{Decohered maps}
	
	We now focus on a subfamily of CP maps that can be used to represent classical processes as a subset of quantum ones. This move from quantum to classical is made by the decoherence map.
	
	\begin{definition}[Decoherence]
		The \textbf{decoherence} map $\delta_n : n\to n$ is defined as: $\delta_n :\rho \mapsto \sum\limits_{x\in \{0,1\}^n } \dyad{x}{x}\rho\dyad{x}{x}$.
	\end{definition}
	
	The decoherence map measures a register of qubits in the computational basis and returns a diagonal density matrix whose diagonal coefficients are the probabilities of observing the different binary words. This amounts to keeping only the diagonal elements of a density matrix. The decoherence map is idempotent, $\delta^2 =\delta $, it is a projector on the commutative algebra of diagonal matrices. We have $\delta_0 = id_0 $ and $\delta_n \otimes \delta_m = \delta_{n+m}$. Classical processes are obtained by measuring a register of qubits, applying a quantum process, and then measuring again.
	
	\begin{definition}[Decohered map]
		The \textbf{decohered} version of a CP map $f:n\to m $ is $\dec{f}= \delta_m \circ f\circ \delta_n $. A CP map is said to be \textbf{decohered} if $\dec{f}=f$.
	\end{definition}
	
	Decohering is an idempotent operation $(f^\circ)^\circ = f^\circ$. We have $\dec{id_0}= id_0 $ and $\dec{(f\otimes g)}=f^\circ \otimes g^\circ $ but not $(f\circ g)^\circ = f^\circ \circ g^\circ$ in general. However, we have a weaker property: $(f^\circ \circ g)^\circ = (f \circ g^\circ )^\circ = f^\circ \circ g^\circ$. This implies that decohered maps are stable by composition and tensor product. Decoherence acts as the identity for decohered maps, $id_{n}^\circ = \delta_n $ and $\delta_m \circ f^\circ = f^\circ \circ \delta_n = f^\circ $. The discard and the decoherence maps are decohered, indeed $\disc^\circ = \disc$ and $\delta_{n}^\circ = \delta_n$. To fully characterize decohered maps we define the following embedding of matrices with positive real coefficients into CP maps. 
	
	\begin{definition}
		Given $A\in \mathcal{M}_{2^m \times 2^n }(\mathbb{R}_+)$ we define the CP map $\uwave{A}: n\to m $ as: $\uwave{A}: \rho \mapsto\sum_{x,y} A_{y,x} \dyad{y}{x}\rho\dyad{x}{y}$.
	\end{definition}
	
	In this paper we write $\mathbb{R}_+ $ for the positive real numbers including $0$, and $\mathbb{R}_{+}^* $ for the strictly positive real numbers, excluding zero. The decohered pure maps are linked to real matrices as follows.
	
	\begin{proposition}\label{prop:decpure}
		$\underline{A}^\circ = \uwave{|A|^2}$ where $\left(|A|^2 \right)_{y,x} = |A_{y,x}|^2 $.
	\end{proposition}
	
	Notice that the condition of $A$ being an isometry translates to $|A|^2 $ being stochastic, as expected. 	
	Notice also from the above, that for any $A\in \mathcal{M}_{2^m \times 2^n }(\mathbb{R}_+)$, $\uwave{A}= \underline{\sqrt A}^\circ$ where $\sqrt -$ is the entry-wise square-root operation on matrices. This embedding is even compositional:
	
	\begin{proposition}\label{prop:wavecomp}
		$\uwave{I_n}= \delta_n $, $\uwave{A}\otimes \uwave{B}= \uwave{A\otimes B}$, and $\uwave{A}\circ \uwave{B}= \uwave{AB}$.
	\end{proposition}
	
	Looking at a specific example, given a vector $v$, $\uwave{v}$ is a diagonal matrix with the coefficients of $v$ on the diagonal. In particular, if $v$ encodes a probability distribution we get a diagonal density matrix. As seen in Proposition \ref{prop:wavecomp}, the decoherence map is just the embedding of the identity $\uwave{I_{n}}= \delta_n $. Interestingly, the discard map is also in the image of this embedding, let $\textbf{1}_{n} $ be the full of $1$ row matrix of size $2^n$ we have for any $\rho$: $\uwave{\textbf{1}_{n}}(\rho)= \sum_{x} \bra{x}\rho\ket{x}= \tr(\rho)= \disc_n (\rho)$. From this, it follows that decohered maps are exactly the image of the embedding. Indeed, given a CP map $f:n\to m$ with purification $A\in \mathcal{M}_{2^{m+k} \times 2^n }(\mathbb{C})$, we can compute the corresponding real matrix as:
	
	\begin{align*}
		f^\circ 
		&= ((\disc_k \otimes id_m )\circ \underline{A})^\circ
		= \delta_m \circ (\disc_k \otimes id_m )\circ \underline{A}\circ \delta_n
		= (\disc_k \otimes \delta_m ) \circ \underline{A}\circ \delta_n
		= (\disc_k \otimes \delta_m ) \circ \delta_{k+m}\circ \underline{A}\circ \delta_n\\
		&= (\disc_k \otimes \delta_m ) \circ \underline{A}^\circ
		= (\uwave{\textbf{1}_{k}} \otimes \uwave{I_{m}} ) \circ \uwave{|A|^2 }
		= \uwave{(\textbf{1}_{k} \otimes I_{m})|A|^2 }
	\end{align*}
	
	This provides us with a complete understanding of decohered maps $n\to m$, they can be identified with matrices in $\mathcal{M}_{2^m \times 2^n }(\mathbb{R}_+ )$, and form their composition and tensor product in the same way.
	
	The main takeaway from this section is that there are two distinct ways to embed some matrices into CP maps. The first is the embedding of complex matrices as pure maps. The second one is the embedding of matrices with positive real coefficients as decohered maps.
	
	From a diagrammatical point of view, complex matrices are depicted by the original ZX-calculus. Thus, the corresponding pure maps have natural diagrammatic representations as a fragment of the discard ZX-calculus. The aim of this paper is exactly to figure out a natural fragment of discard ZX-calculus able to depict matrices with real coefficients and their embedding as decohered maps.
	\begin{wrapfigure}{r}{0.4\textwidth}\label{fig4}
		\centering\input{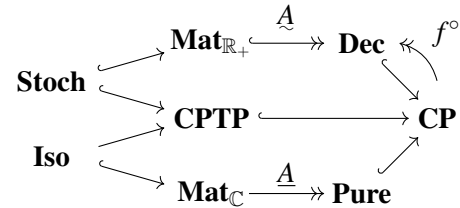}
		\caption{Diagram of inclusions. Double head means surjective and a hook means injective.}
	\end{wrapfigure}
	We end this section by reformulating the situation so far in the language of props. Let $\textbf{Mat}_{\mathbb{C}}$ be the prop whose arrows $n\to m $ are matrices in $\mathcal{M}_{2^m \times 2^n }(\mathbb{C})$, and \textbf{Iso} the subprop of isometries. Let $\textbf{Mat}_{\mathbb{R_+ }}$ be the prop whose arrows $n\to m $ are matrices in $\mathcal{M}_{2^m \times 2^n }(\mathbb{R_ + })$ and \textbf{Stoch} the subprop of stochastic matrices. Let $\textbf{CP}$ be the prop whose arrows are CP maps $n\to m $, and \textbf{CPTP} and \textbf{Pure} respectively the sub prop of trace-preserving and pure CP maps. We can also define a prop \textbf{Dec} of decohered maps, but one needs to be careful, as this is not a subprop of \textbf{CP} as the role of the identity is played by $\delta_n$. We have the commutative diagram of Figure \hyperref[fig4]{1}. In this diagram, all arrows are prop morphisms except for $f\mapsto f^\circ $ which doesn't preserve composition and the embeddings of \textbf{Dec} into \textbf{CP} and \textbf{Stoch} into \textbf{CPTP} that do not preserve the identities. We see that decohered maps can be obtained either by decohering CP maps or by embedding matrices with positive real coefficients.
	
	\section{Discard ZX-calculus}\label{sec:disczx}
	
	The discard ZX-calculus has been introduced in the form we present here and proven complete in \cite{carette2021completeness}. We only change the normalization in order to have normalized states, simplify the trigonometric formulas, and make the scalars explicit.
	
	\subsection{Discard ZX-diagrams}
	
	Discard ZX-diagrams with $n$ inputs and $m$ outputs $D:n\to m $ are inductively defined as:
	
	\begin{itemize}
		\item The empty diagram, a wire, a swap, a cup, or a cap:
		\begin{center}
			$\begin{tikzpicture}
\end{tikzpicture}
:0\to 0 \qquad \begin{tikzpicture}
	\begin{pgfonlayer}{nodelayer}
		\node [style=none] (0) at (-0.5, 0) {};
		\node [style=none] (1) at (0.5, 0) {};
	\end{pgfonlayer}
	\begin{pgfonlayer}{edgelayer}
		\draw (0.center) to (1.center);
	\end{pgfonlayer}
\end{tikzpicture}
:1\to 1 \qquad \input{swap.tikz}: 2\to 2 \qquad \begin{tikzpicture}
	\begin{pgfonlayer}{nodelayer}
		\node [style=none] (0) at (0.25, -0.5) {};
		\node [style=none] (2) at (0.25, 0.5) {};
		\node [style=none] (4) at (-0.25, 0) {};
	\end{pgfonlayer}
	\begin{pgfonlayer}{edgelayer}
		\draw [in=90, out=-180] (2.center) to (4.center);
		\draw [in=180, out=-90] (4.center) to (0.center);
	\end{pgfonlayer}
\end{tikzpicture}
: 0\to 2 \qquad \begin{tikzpicture}
	\begin{pgfonlayer}{nodelayer}
		\node [style=none] (0) at (0, -0.5) {};
		\node [style=none] (2) at (0, 0.5) {};
		\node [style=none] (4) at (0.5, 0) {};
	\end{pgfonlayer}
	\begin{pgfonlayer}{edgelayer}
		\draw [in=90, out=0] (2.center) to (4.center);
		\draw [in=0, out=-90] (4.center) to (0.center);
	\end{pgfonlayer}
\end{tikzpicture}
: 2\to 0$
		\end{center}
		
		\item One of the discard ZX generators, where $\alpha$ is a parameter in $[0,2\pi[$ and $s\in \mathbb{R}_+ $:
		
		\begin{center}
			$ \input{gspider.tikz}: n\to m \qquad \input{rspider.tikz}:n\to m \qquad \begin{tikzpicture}
	\begin{pgfonlayer}{nodelayer}
		\node [style=hadamard] (0) at (0, 0) {};
		\node [style=none] (1) at (-1, 0) {};
		\node [style=none] (2) at (1, 0) {};
	\end{pgfonlayer}
	\begin{pgfonlayer}{edgelayer}
		\draw (1.center) to (0);
		\draw (0) to (2.center);
	\end{pgfonlayer}
\end{tikzpicture}
:1\to 1 \qquad \begin{tikzpicture}
	\begin{pgfonlayer}{nodelayer}
		\node [style=scalar] (5) at (0, 0) {$s$};
	\end{pgfonlayer}
\end{tikzpicture}
: 0\to 0 \qquad\input{disc.tikz}:1\to 0$
		\end{center}
		
		\item A \textbf{parallel composition} of two diagrams $D_1 : a\to b $ and $D_2 : c\to d$: $\quad\input{par.tikz}$
		\vspace{0.25cm}
		\item A \textbf{sequential composition} of two diagrams $D_1 : a\to b $ and $D_2 : b\to c$: $\quad\input{seq.tikz}$
	\end{itemize}
	
	The parallel composition is associative and the empty diagram behaves as a neutral element. This allows us to define without ambiguity the identity diagram $id_n : n\to n $ as a parallel composition of $n$ wires. Notice that $id_0$ is the empty diagram, and $id_1 $ is a wire. We depict the identity diagram as $\input{id.tikz}: n\to n $. The sequential composition is associative and the identity diagram behaves as a neutral element. Parallel and sequential composition commute.
	
	\begin{center}
		\scalebox{0.7}{\input{straxioms.tikz}}
	\end{center}
	
	The axioms stated so far ensure that all ways to decompose a diagram as sequential and parallel compositions of generators are equivalent, thus we can without ambiguity drop the grey boxes used to keep track of how the diagram was constructed. In the language of categories, we exactly stated in diagrammatical form the defining axioms of a pro. The swap, the cup, and the cap generators satisfy for all diagram $D:n\to m$:
	
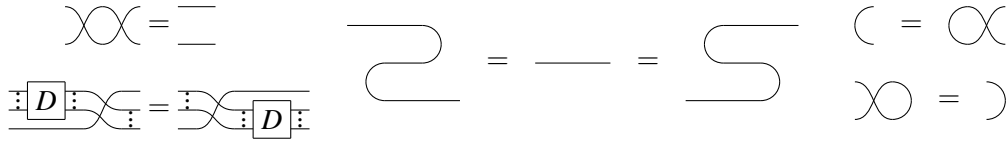
	\begin{center}
		\input{swapsnakeaxioms.tikz}
	\end{center}
	
	The reader familiar with category theory and string diagrams will recognize here the definition of the free compact prop spanned by the set of discard ZX generators. All generators of the discard ZX-calculus are flexsymmetric, i.e.~we have the equations:
	\begin{center}
		\input{flexsym.tikz}
	\end{center}
	where $\sigma$ can be any permutation of wires encoded by a combination of swaps. So the only information that is needed to specify a diagram is its underlying graph. All the axioms given so far are topological in nature and correspond to intuitive deformations of diagrams. They are shared by most of the similar diagrammatical languages. Later we will define the decohered ZX diagrams in the same way, just replacing the discard ZX generators with the decohered ZX generators.
	\subsection{Rules}
	
	To complete our definition of the discard ZX-calculus we add the following rewriting rules:
	
	\begin{figure}[H]\label{fig1}
		\centering
		\input{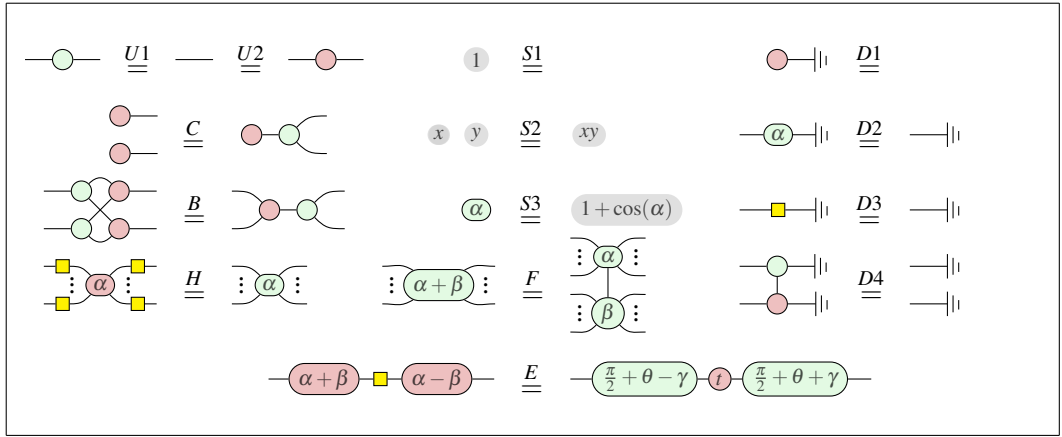}
		\caption{Rules of the discard ZX-calculus.}
	\end{figure}
	Where: $t=\arccos(\cos^2 (\beta) -\cos^2 (\alpha))$, $\gamma=\arg(\cos(\alpha) +i\sin(\beta))$ and $\theta=\arg(\cos(\beta) +i\sin(\alpha))$. Notice that as a convention we do not write the phase inside a red or green node when $\alpha=0$.
	\subsection{Interpretation}
	
	To each discard ZX-diagram $D: n\to m $ we assign a unique CP map $\interp{D}:n\to m$ inductively using:
	
	\begin{center}
		$\interp{\input{par.tikz}}= \interp{D_1}\otimes \interp{D_2} \qquad \qquad\interp{\input{seq.tikz}}= \interp{D_2} \circ \interp{D_1}$
	\end{center}
	
	The elementary diagrams $D:n\to m$ are interpreted as follows:
	
	\begin{center}
		$\interp{\input{id.tikz}} = \underline{I_{2^n}} \qquad \interp{\input{swap.tikz}} = \underline{\sum_{x,y} \ket{xy}\bra{yx}}\qquad\interp{\begin{tikzpicture}
	\begin{pgfonlayer}{nodelayer}
		\node [style=none] (0) at (0.25, -0.5) {};
		\node [style=none] (2) at (0.25, 0.5) {};
		\node [style=none] (4) at (-0.25, 0) {};
	\end{pgfonlayer}
	\begin{pgfonlayer}{edgelayer}
		\draw [in=90, out=-180] (2.center) to (4.center);
		\draw [in=180, out=-90] (4.center) to (0.center);
	\end{pgfonlayer}
\end{tikzpicture}
}= \frac{1}{2}\underline{\ket{00} + \ket{11}}\qquad\interp{\begin{tikzpicture}
	\begin{pgfonlayer}{nodelayer}
		\node [style=none] (0) at (0, -0.5) {};
		\node [style=none] (2) at (0, 0.5) {};
		\node [style=none] (4) at (0.5, 0) {};
	\end{pgfonlayer}
	\begin{pgfonlayer}{edgelayer}
		\draw [in=90, out=0] (2.center) to (4.center);
		\draw [in=0, out=-90] (4.center) to (0.center);
	\end{pgfonlayer}
\end{tikzpicture}
} =2\underline{\bra{00} + \bra{11}}$\\
		\vspace{0.5cm}
		$\interp{\input{gspider.tikz}} = 2^{n-1}\underline{\ket{0}^{\otimes m} \bra{0}^{\otimes n} + e^{i\alpha}\ket{1}^{\otimes m}\bra{1}^{\otimes n}}\qquad \interp{\input{rspider.tikz}} = 2^{n-1}\underline{\ket{+}^{\otimes m} \bra{+}^{\otimes n} + e^{i\alpha}\ket{-}^{\otimes m}\bra{-}^{\otimes n}}$\\
		\vspace{0.5cm}
		$\interp{\begin{tikzpicture}
	\begin{pgfonlayer}{nodelayer}
		\node [style=hadamard] (0) at (0, 0) {};
		\node [style=none] (1) at (-1, 0) {};
		\node [style=none] (2) at (1, 0) {};
	\end{pgfonlayer}
	\begin{pgfonlayer}{edgelayer}
		\draw (1.center) to (0);
		\draw (0) to (2.center);
	\end{pgfonlayer}
\end{tikzpicture}
}= \underline{\ket{+}\bra{0}+ \ket{-}\bra{1}}\qquad\interp{\begin{tikzpicture}
	\begin{pgfonlayer}{nodelayer}
		\node [style=scalar] (5) at (0, 0) {$s$};
	\end{pgfonlayer}
\end{tikzpicture}
} = s \qquad \interp{\input{disc.tikz}}=\disc_1 $\\
	\end{center}
	
	One can check that all equivalences of diagrams stated so far are sound with respect to this interpretation, ensuring that the obtained CP map is unique. The discard ZX-calculus can represent all CP maps.
	
	\begin{proposition}[Universality]\label{prop:univcpm}
		For any CP map $f: n\to m $, there is a discard $ZX$ diagram $D: n\to m$ such that $\interp{D}=f$.
	\end{proposition}
	
	Furthermore, the rules of the discard calculus are powerful enough to show any equality between CP maps.
	
	\begin{proposition}[Completeness]\label{prop:compcpm}
		Given two ZX-diagrams $D_1 : n\to m $ and $D_2 : n\to m $ such that $\interp{D_1 }= \interp{D_2 }$ then $D_1 = D_2 $ modulo the rules of Figure \hyperref[fig1]{2}.
	\end{proposition}
	
	\subsection{Picturing decoherence}
	
	Discard ZX-calculus is expressive and powerful enough to handle any equational reasoning on CP maps; hence we should be able to picture decohered maps in this formalism. Indeed, the decoherence map and the decohered maps are depicted as: $\interp{\scalebox{0.8}{\input{dec.tikz}}}=\delta_1$ and $\interp{\scalebox{0.8}{\input{decD.tikz}}}=\interp{D}^\circ$, where $D$ is any discard ZX-diagram. We could carry all the study of decohered maps in this way through the discard ZX-calculus, using, but our goal here is to provide decohered maps with a ZX-calculus of their own.
	
	\section{Decohered ZX-calculus}\label{sec:deczx}
	
	In this section, we will define a decohered ZX-calculus able to depict a restricted family of decohered maps. To avoid the confusion between the decohered and the discard ZX we will use grey wires for the decohered version, conveying the idea that those wires are intrinsically classical.
	
	\subsection{The decohered generators}
	
	We start by investigating the decohered maps associated with the generators of the discard ZX-calculus.  For generators with pure interpretation, we can use Proposition \ref{prop:decpure} to get:
	
	\begin{center}
		$\interp{\input{id.tikz}}^\circ = \uwave{I_{2^n}} \qquad \interp{\input{swap.tikz}}^\circ = \uwave{\sum_{x,y} \ket{xy}\bra{yx}}\qquad\interp{\begin{tikzpicture}
	\begin{pgfonlayer}{nodelayer}
		\node [style=none] (0) at (0.25, -0.5) {};
		\node [style=none] (2) at (0.25, 0.5) {};
		\node [style=none] (4) at (-0.25, 0) {};
	\end{pgfonlayer}
	\begin{pgfonlayer}{edgelayer}
		\draw [in=90, out=-180] (2.center) to (4.center);
		\draw [in=180, out=-90] (4.center) to (0.center);
	\end{pgfonlayer}
\end{tikzpicture}
}^\circ= \frac{1}{2}\uwave{\ket{00} + \ket{11}}\qquad\interp{\begin{tikzpicture}
	\begin{pgfonlayer}{nodelayer}
		\node [style=none] (0) at (0, -0.5) {};
		\node [style=none] (2) at (0, 0.5) {};
		\node [style=none] (4) at (0.5, 0) {};
	\end{pgfonlayer}
	\begin{pgfonlayer}{edgelayer}
		\draw [in=90, out=0] (2.center) to (4.center);
		\draw [in=0, out=-90] (4.center) to (0.center);
	\end{pgfonlayer}
\end{tikzpicture}
}^\circ =2\uwave{\bra{00} + \bra{11}}$\\
		\vspace{0.5cm}
		$\interp{\input{gspider.tikz}}^\circ = 2^{n-1}\uwave{\ket{0}^{\otimes m} \bra{0}^{\otimes n} + \ket{1}^{\otimes m}\bra{1}^{\otimes n}}\qquad \interp{\begin{tikzpicture}
	\begin{pgfonlayer}{nodelayer}
		\node [style=scalar] (5) at (0, 0) {$s$};
	\end{pgfonlayer}
\end{tikzpicture}
}^\circ = s$\\
	\end{center}
	
	Notice that the green spiders lose their parameters. For the Hadamard gate and the discard map, we get: $\interp{\begin{tikzpicture}
	\begin{pgfonlayer}{nodelayer}
		\node [style=hadamard] (0) at (0, 0) {};
		\node [style=none] (1) at (-1, 0) {};
		\node [style=none] (2) at (1, 0) {};
	\end{pgfonlayer}
	\begin{pgfonlayer}{edgelayer}
		\draw (1.center) to (0);
		\draw (0) to (2.center);
	\end{pgfonlayer}
\end{tikzpicture}
}^\circ = \frac{1}{2}\underline{\begin{pmatrix}
			1 & 1 \\ 1 & -1
	\end{pmatrix}}^\circ =\frac{1}{2}\uwave{\begin{pmatrix}
			1 & 1 \\ 1 & 1
	\end{pmatrix}}= \interp{\begin{tikzpicture}
	\begin{pgfonlayer}{nodelayer}
		\node [style=vert] (0) at (-0.5, 0) {};
		\node [style=none] (3) at (-1.5, 0) {};
		\node [style=vert] (4) at (0.5, 0) {};
		\node [style=none] (5) at (1.5, 0) {};
	\end{pgfonlayer}
	\begin{pgfonlayer}{edgelayer}
		\draw (3.center) to (0);
		\draw (5.center) to (4);
	\end{pgfonlayer}
\end{tikzpicture}
}^\circ$ and $\interp{\input{disc.tikz}}^\circ = \uwave{\begin{pmatrix}
			1 & 1
	\end{pmatrix}}= \interp{\begin{tikzpicture}
	\begin{pgfonlayer}{nodelayer}
		\node [style=vert] (0) at (0.5, 0) {};
		\node [style=none] (3) at (-0.5, 0) {};
	\end{pgfonlayer}
	\begin{pgfonlayer}{edgelayer}
		\draw (3.center) to (0);
	\end{pgfonlayer}
\end{tikzpicture}
}^\circ$. So we see that in the decohered ZX-calculus the Hadamard gate and the discard will not be useful as generators, their role being already fulfilled by the decohered green spider. It remains to consider the case of the red spider: $\interp{\input{rspider.tikz}}^\circ = 2^{1-m}\left(\underline{\sum_{x,y} \frac{1+(-1)^{|xy|}e^{i\alpha}}{2} \dyad{x}{y}}\right)^\circ=2^{1-m} \uwave{\sum_{x,y} \frac{1+(-1)^{|xy|}\cos(\alpha)}{2} \dyad{x}{y}}$, where $|xy|$ is the number of ones in the concatenation of $x$ and $y$. We see an asymmetry between green and red spiders, here the parameter is preserved but takes a different form. We will choose a different parametrization for the decohered red spider, by setting $p=\frac{1-\cos(\alpha)}{2}$ we have $p \in [0,1]$. It follows that: $\interp{\begin{tikzpicture}
	\begin{pgfonlayer}{nodelayer}
		\node [style=rouge] (454) at (0, 0) {$\alpha$};
		\node [style=none] (457) at (1, 0) {};
	\end{pgfonlayer}
	\begin{pgfonlayer}{edgelayer}
		\draw (457.center) to (454);
	\end{pgfonlayer}
\end{tikzpicture}
}^\circ = \uwave{\begin{pmatrix}
			\frac{1+\cos(\alpha)}{2} \\ \frac{1-\cos(\alpha)}{2}
	\end{pmatrix}}= \uwave{\begin{pmatrix}
			1-p \\ p
	\end{pmatrix}}= \begin{pmatrix}
		1-p & 0 \\0 & p
	\end{pmatrix}$. We get the density matrix representation of a biased coin giving $\ket{0}$ with probability $1-p$ and $\ket{1}$ with probability $p$. In particular, when $p=0$ we get $\dyad{0}{0}$ and when $p=1$ we get $\dyad{1}{1}$. From decohered red states with $\alpha \neq \pi$ we can recover a way to parametrize decohered green spiders:
	
	\[\interp{\begin{tikzpicture}
	\begin{pgfonlayer}{nodelayer}
		\node [style=rouge] (454) at (0, 0) {$\alpha$};
		\node [style=none] (457) at (1, 0) {};
		\node [style=scalar] (458) at (-2, 0) {$\frac{2}{1+\cos(\alpha)}$};
	\end{pgfonlayer}
	\begin{pgfonlayer}{edgelayer}
		\draw (457.center) to (454);
	\end{pgfonlayer}
\end{tikzpicture}
}^\circ = \frac{2}{1+\cos(\alpha)}\uwave{\begin{pmatrix}
			\frac{1+\cos(\alpha)}{2} \\ \frac{1-\cos(\alpha)}{2}
	\end{pmatrix}}= \uwave{\begin{pmatrix}
			1 \\ \frac{1-\cos(\alpha)}{1+\cos(\alpha)}
	\end{pmatrix}}= \uwave{\begin{pmatrix}
			1 \\ \frac{p}{1-p}
	\end{pmatrix}}=\uwave{\begin{pmatrix}
			1 \\ \mu
	\end{pmatrix}}\]
	
	Where $\mu= \frac{p}{1-p} \in \mathbb{R}_+ $. Then we can define a parametrized decohered green spider as:
	
	\[\interp{\input{paradecgn.tikz}}^\circ = 2^{n-1} \uwave{\ket{0}^{\otimes m} \bra{0}^{\otimes n} + \mu\ket{1}^{\otimes m}\bra{1}^{\otimes n}} \]
	
	We are now ready to define the generators of the decohered ZX-calculus, from now on we provide the interpretations of decohered ZX-diagrams directly in $\textbf{Mat}_{\mathbb{R}_+ }$. 
	
	\begin{center}
		$\interp{\input{did.tikz}} = I_{2^n} \quad \interp{\input{dswap.tikz}} = \sum_{x,y} \ket{xy}\bra{yx}\quad\interp{\begin{tikzpicture}
	\begin{pgfonlayer}{nodelayer}
		\node [style=none] (0) at (0.25, -0.5) {};
		\node [style=none] (2) at (0.25, 0.5) {};
		\node [style=none] (4) at (-0.25, 0) {};
	\end{pgfonlayer}
	\begin{pgfonlayer}{edgelayer}
		\draw [style=dgrey, in=90, out=-180] (2.center) to (4.center);
		\draw [style=dgrey, in=180, out=-90] (4.center) to (0.center);
	\end{pgfonlayer}
\end{tikzpicture}
}= \frac{1}{2}(\ket{00} + \ket{11})\quad\interp{\begin{tikzpicture}
	\begin{pgfonlayer}{nodelayer}
		\node [style=none] (0) at (0, -0.5) {};
		\node [style=none] (2) at (0, 0.5) {};
		\node [style=none] (4) at (0.5, 0) {};
	\end{pgfonlayer}
	\begin{pgfonlayer}{edgelayer}
		\draw [style=dgrey, in=90, out=0] (2.center) to (4.center);
		\draw [style=dgrey, in=0, out=-90] (4.center) to (0.center);
	\end{pgfonlayer}
\end{tikzpicture}
} =2(\bra{00} + \bra{11}) \quad \interp{\begin{tikzpicture}
	\begin{pgfonlayer}{nodelayer}
		\node [style=scalar] (5) at (0, 0) {$s$};
	\end{pgfonlayer}
\end{tikzpicture}
} = s$\\
		\vspace{0.5cm}
		$\interp{\input{dgspider.tikz}} = 2^{n-1}(\ket{0}^{\otimes m} \bra{0}^{\otimes n} + \mu\ket{1}^{\otimes m}\bra{1}^{\otimes n})\qquad \interp{\input{drspider.tikz}} =2^{1-m} \sum_{x,y} \left(\begin{cases}
			1-p \text{ if } $|xy|=0$ \\
			p \text{ if } $|xy|=1$
		\end{cases}\right)\dyad{x}{y}$\\
	\end{center}
	
	The parameters of red spiders are probabilities $p\in[0,1]$. The parameters of green spiders are positive real numbers $\mu\in \mathbb{R}_+$. As a convention, we don't write anything in decohered green spiders when $\mu= 1 $ and in decohered red spiders when $p=0$. Those generators can be embedded in the discard ZX-calculus as: \input{dectodisc1.tikz} and \input{dectodisc2.tikz}, where respectively $\alpha = \arccos\left(1-2\frac\mu{1+\mu}\right)$ and $\beta = \arccos(1-2p)$.
	
	\subsection{Rules}
	
	The wires and the nodes in the decohered ZX-diagrams behave in the same way as in the discard ZX-calculus, meaning that here also diagrams can be seen as graphs. However, the core set of rules is different.
	
	\begin{figure}[H]\label{fig2}
		\centering
		\input{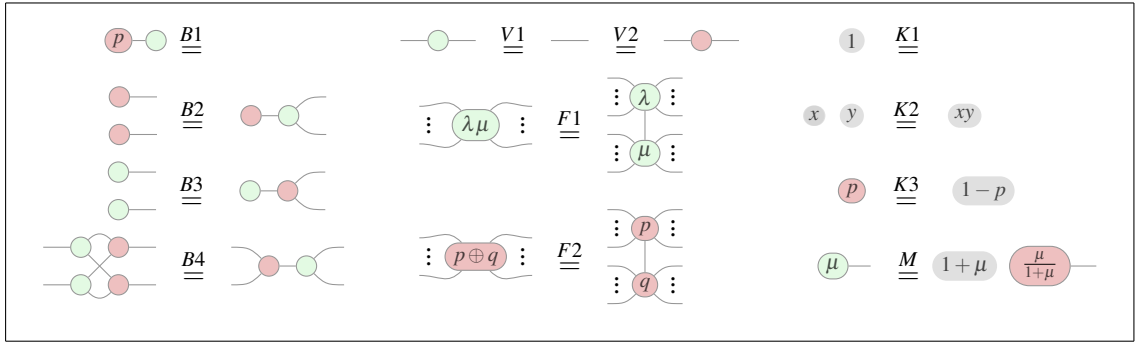}
		\caption{Rules of the decohered ZX-calculus (with one rule missing)}
	\end{figure}
	
	There is one additional rule necessary to make the language complete, we postpone its definition to Section \ref{sec:unicomp}, where we will have the necessary tools to state it clearly. All those rules are sound for the interpretation as matrices with positive real coefficients. The commutative operation $\oplus$ generalizes the XOR to the whole unit interval: $p\oplus q = p + q - 2pq $. $(.\oplus., 0)$ forms a commutative monoid: $p\oplus(q\oplus r) = (p\oplus q)\oplus r$ and $p \oplus 0 = p $. Moreover, $\frac12$ is an absorbing element: $p\oplus \frac{1}{2} = \frac{1}{2}$, and $.\oplus1$ performs a symmetry around this absorbing element: $p \oplus 1 = 1-p $. The fact that decohered red spiders fuse according to the $\oplus$ operation is remarkable, in particular, this provides a nice probabilistic interpretation of decohered red spiders with small arity.  Decohered red spiders $0\to 1 $ allow to express any probability distribution over $\ket{0}$ and $\ket{1}$: $\interp{\begin{tikzpicture}
	\begin{pgfonlayer}{nodelayer}
		\node [style=drouge] (0) at (-2, 0) {};
		\node [style=none] (1) at (-1, 0) {};
	\end{pgfonlayer}
	\begin{pgfonlayer}{edgelayer}
		\draw [style=dgrey] (1.center) to (0);
	\end{pgfonlayer}
\end{tikzpicture}
}= \ket{0}$, $\interp{\begin{tikzpicture}
	\begin{pgfonlayer}{nodelayer}
		\node [style=drouge] (0) at (-2, 0) {$1$};
		\node [style=none] (1) at (-1, 0) {};
	\end{pgfonlayer}
	\begin{pgfonlayer}{edgelayer}
		\draw [style=dgrey] (1.center) to (0);
	\end{pgfonlayer}
\end{tikzpicture}
}= \ket{1}$ and $\interp{\begin{tikzpicture}
	\begin{pgfonlayer}{nodelayer}
		\node [style=drouge] (0) at (0, 0) {$p$};
		\node [style=none] (1) at (1, 0) {};
	\end{pgfonlayer}
	\begin{pgfonlayer}{edgelayer}
		\draw [style=dgrey] (0) to (1.center);
	\end{pgfonlayer}
\end{tikzpicture}
}= \begin{pmatrix}
		1-p \\ p
	\end{pmatrix}$. The decohered red spiders $1\to 1$ flip the input bit with probability $p$, in particular, when $p=1$ we get the NOT gate: $\interp{\begin{tikzpicture}
	\begin{pgfonlayer}{nodelayer}
		\node [style=drouge] (0) at (0, 0) {$p$};
		\node [style=none] (1) at (1, 0) {};
		\node [style=none] (2) at (-1, 0) {};
	\end{pgfonlayer}
	\begin{pgfonlayer}{edgelayer}
		\draw [style=dgrey] (0) to (1.center);
		\draw [style=dgrey] (2.center) to (0);
	\end{pgfonlayer}
\end{tikzpicture}
}=\scalebox{0.8}{$\begin{pmatrix}
			1-p & p \\
			p & 1-p
		\end{pmatrix}$}$ and $\interp{\begin{tikzpicture}
	\begin{pgfonlayer}{nodelayer}
		\node [style=none] (1) at (1, 0) {};
		\node [style=drouge] (2) at (0, 0) {$1$};
		\node [style=none] (3) at (-1, 0) {};
	\end{pgfonlayer}
	\begin{pgfonlayer}{edgelayer}
		\draw [style=dgrey] (3.center) to (2);
		\draw [style=dgrey] (1.center) to (2);
	\end{pgfonlayer}
\end{tikzpicture}
}= \scalebox{0.8}{$\begin{pmatrix}
			0 & 1 \\
			1 & 0
		\end{pmatrix}$}$. Rule \hyperref[rule:M]{M} implies in particular: \input{coin.tikz}. 
	%
	The reader familiar with the ZX-calculus can see that trying to prove the soundness of those rules through the translation to the discard ZX-calculus can be very difficult, however, the exercise provides interesting insights. For example Rules \hyperref[rule:B1]{B1}, \hyperref[rule:B2]{B2}, \hyperref[rule:B3]{B3} and \hyperref[rule:B4]{B4} rely crucially on Rule \hyperref[rule:D4]{D4} of the discard ZX-calculus. Rule \hyperref[rule:M]{M} in the case $\mu=0$ is the decohered counterpart of a well-known equality between the red state with parameter $\frac{\pi}{2}$ and the green state with parameter $\frac{-\pi}{2}$. Deriving Rule \hyperref[rule:F2]{F2} in the discard ZX-calculus, though theoretically possible thanks to completeness, remains an open problem.

	\section{Universality and Completeness}\label{sec:unicomp}
	
	Now that the decohered ZX-calculus is fully defined we can focus on its universality and completeness. As we will see, the decohered ZX-calculus is not able to express all decohered maps, however, it is still universal for an interesting subset of them. The set of rules given in Figure \hyperref[fig3]{4} is complete. Proving this fact will require the introduction of several concepts. In particular, the notion of support of a matrix will be very important.
	
	\begin{definition}[Support]
		The \textbf{support} of a matrix $A\in \mathcal{M}_{2^m \times 2^n}(\mathbb{R}_+ )$ is $\mathrm{supp}(A)=\{(x,y)\in \mathbb{F}_{2}^m \times \mathbb{F}_{2}^n  ~|~ A_{x,y}\neq 0 \}$. We say that a matrix $A$ has \textbf{full support} when $\mathrm{supp}(A)=\mathbb{F}_{2}^{m+n}$, and that is has \textbf{affine support} when $\mathrm{supp}(A)$ is an affine subspace of $\mathbb{F}_{2}^{m+n}$.
	\end{definition}
	
	\subsection{Scalable notations}
	
	To simplify the depiction of arbitrary large diagrams we will adapt the notations introduced in \cite{carette2019szx} and use thick wires to depict registers of $n$ qubits, in other words, $n$ wires side by side. To divide or gather wires when needed we introduce the following notations that are inverse of each other:
	
	\begin{center}
		\input{scaleaxioms.tikz}
	\end{center}
	
	Those notations are useful to define large diagrams inductively. We generalize the green and red decohered spider to wires of size $k$ as:
	
	\begin{center}
		\input{scalespiderdef.tikz}
	\end{center}
	
	Notice that now decohered red spiders are parametrized by a vector of probabilities $\mathbf{p}\in [0,1]^{k}$ which is the concatenation of $\mathbf{p_1}$ and $\textbf{q}\in [0,1]^{k-1}$. In the same way, decohered green spiders are parametrized by a vector of positive real numbers $\mathbf{\mu}\in \mathbb{R}_{+}^{k}$ which is the concatenation of $\mathbf{\mu_1}$ and $\mathbf{\lambda}\in \mathbb{R}_{+}^{k-1}$. We keep the same convention as before and do not indicate the parameters when it is a vector full of $O$, in the red case, or full of $1$, in the green case. We inductively define the \textbf{matrix arrows} indexed by a matrix $A\in \mathcal{M}_{m\times n }(\mathbb{F}_2 )$ as:
	\begin{center}
		\input{matdef.tikz}
	\end{center}
	\begin{proposition}\label{prop:matrules}
		Matrix arrows satisfy:
		\begin{center}
			\input{matrules2.tikz}
		\end{center}
	\end{proposition}
	For simplicity we use an informal presentation here, we let the interested reader consult \cite{carette2019szx} for a formal approach and full proofs of those properties. A straightforward induction shows that scaled spiders satisfy the same equation as size one spiders, the $\oplus$ and product operations being component-wise for \hyperref[rule:sF1]{sF1} and \hyperref[rule:sF2]{sF2}.
	\begin{figure}[H]\label{fig3}
		\centering
		\input{scaledeczxaxioms.tikz}
		\caption{Scaling of the rules of the decohered ZX-calculus}
	\end{figure}
	Where $\mathbf{1}$ is the column matrix full of $1$ of size $n$, $\lambda \in \mathbb{R}_{+}^*$, $\mathbf{\mu} \in (\mathbb{R}_{+}^*)^{n}$ and $\mathfrak{s}_n \in \mathcal{M}_{n\times (2^n -1) }(\mathbb{F}_2 )$ is the \textbf{subset matrix} defined as $(\mathfrak{s}_{n})_{i,x}=\delta_{i\in x}= x_i$, with $1\leq i\leq n$ (interpreted as elements in a set of size $n$) and $x\in \mathbb{F}_{2}^{n}\setminus\{0\}$ (interpreted as a non-empty subset of a set with $n$ element). Notice that in Rule \hyperref[rule:sM]{sM}, all operations are coefficient-wise. In Rule \hyperref[rule:L]{L} we have $s=1+\lambda \prod_i \mu_{i}$ and for all $x\in \mathbb{F}_{2}^{n}/\{0\} $, $\nu_x = \prod_{y} \left(\frac{\prod_i \mu_{i}^{x_i} + \lambda \prod_i \mu_{i}^{1-x_i}}{s} \right)^{\frac{-2(-1)^{x\cdot y}}{2^{n}}}$. This complicated rule will make more sense in the framework of diagrammatic Fourier transforms.
	
	\subsection{Diagrammatic Fourier transform}
	
	We recast the work of \cite{kuijpers2019graphical} in the decohered setting. We start by exhibiting a normal form in decohered ZX-calculus able to represent any vector with full support.
	
	\begin{definition}[Fourier normal form]
		A decohered ZX diagram $D: 0 \to n$ is in \textbf{Fourier normal form} if it is of the form: \input{fnf.tikz} where $\mathbf{\lambda} \in (\mathbb{R}_{+}^*)^{2^{n-1}}$ and $\Lambda \in \mathbb{R}_{+}^*$.
	\end{definition}
	
	The interpretation of a Fourier normal form is: $\interp{\input{fnf.tikz}}= \Lambda\sum_{x}\prod_{y\neq 0} \lambda_{y}^{x\cdot y} \ket{x}$. where $x\cdot y = \bigoplus_i x_i y_i $ is the scalar product in $\mathbb{F}_{2}^{n}$. This implies that the interpretation of a Fourier normal form has full support, in fact, we also have the converse.
	
	\begin{proposition}\label{prop:uniquefnf}
		For any vector $v\in \mathbb{R}_{+}^{2^n}$ with full support, there is a unique diagram in Fourier normal form with interpretation $v$. In other words, there is a unique non-zero $\Lambda \in \mathbb{R}_+$ and a unique vector $\lambda \in \mathbb{R}_{+}^{2^n - 1}$ with full support such that: $\interp{\input{fnf.tikz}}=v$. Furthermore, we have $\Lambda= v_0$ and for each $y\in \mathbb{F}_{2}^{n} \setminus \{0\}$: $\lambda_y = \prod_{x} \left(\frac{v_{x}}{v_0}\right)^{\frac{-2(-1)^{x\cdot y}}{2^{n}}}$.
	\end{proposition}
	
	Thus, each vector with full support admits a unique Fourier normal form representation. The subset matrix also admits the following inductive definition: $\input{subdef.tikz}$, where $\textbf{1}$ is the full of $1$ column matrix of size $2^{n}-1$. From it, we can reduce many diagrams to Fourier normal form.
	
	\begin{proposition}\label{prop:lfnf}
		Diagrams in the four following forms reduce to Fourier normal forms assuming that $\lambda$ has full support: $\quad\input{lfnf.tikz}$.
	\end{proposition}
	
	Fourier normal forms behave well under the action of invertible matrix arrows.
	
	\begin{proposition}\label{prop:perm}
		Given any invertible matrix $A\in \mathrm{GL}_n (\mathbb{F}_2)$, we have: $\input{subperm.tikz}$, where $\sigma_{A} \in \mathcal{M}_{2^n \times 2^n }{\mathbb{F}_2}$ is a permutation matrix defined as $(\sigma_{A})_{s,t}=\delta_{At=s}$.
	\end{proposition}
	
	\subsection{Affine diagrams}
	
	We say that a matrix $A\in \mathcal{M}_{2^m \times 2^n } (\mathbb{R}_+ )$ has \textbf{affine support} if $\mathrm{supp}(A)$ is an affine subspace of $\mathbb{F}^{m+n}$. Matrices with affine support have tight links with a particular kind of decohered ZX-diagrams called affine.
	
	\begin{definition}[Affine diagram]
		A diagram of the decohered ZX-calculus is \textbf{affine} if all decohered green spiders in it have parameter $1$, and all decohered red spiders have parameter either $0$ or $1$.
	\end{definition}
	
	Affine diagrams are an avatar of Graphical Affine Algebra (GAA) \cite{bonchi2019graphical}. Most of the results presented here are directly adapted from GAA.
	
	\begin{proposition}\label{prop:sanf}
		An affine diagrams of type $0\to n$ can be rewritten in the form: $\input{sanf.tikz}$.
	\end{proposition}
	
	This allows us to compute the general interpretation of an affine diagram.
	
	\begin{proposition}\label{prop:interpgaa}
		The interpretation of an affine diagram $D:0\to n$ is of the form $\interp{D}=s\sum_{y} \ket{Ay\oplus x}$ with $A\in \mathcal{M}_{n \times k}(\mathbb{F}_2 )$, $x\in \mathbb{F}^n$ and $s\in \mathbb{R}_+ $.
	\end{proposition}
	
	This implies that affine diagrams only depict matrices with affine support. The following result will prove useful in the next section.
	
	\begin{proposition}\label{prop:banf}
		An affine diagram of type $n\to m$ can be rewritten in the form: $\input{banf.tikz}$ where $P$ and $Q$ are invertible.
	\end{proposition}
	
	\subsection{Normal form}
	
	We are now ready to define a general normal form for decohered ZX-diagrams. We first recall the map/state duality which follows from a straightforward use of the snake equation. 
	
	\begin{proposition}[Map/state duality]\label{prop:mapstate}
		To any decohered ZX-diagram $D:n\to m$ we can associate a unique decohered ZX-diagram $\lceil D \rceil: 0 \to n+m$ such that: \input{namedef1.tikz} and \input{namedef2.tikz}. Furthermore, given two diagrams $D_1 , D_2 : n\to m $ we have:
		\begin{center}
			$D_1 = D_2 \Leftrightarrow \lceil D_1 \rceil=\lceil D_2 \rceil \qquad\text{ and }\qquad\interp{D_1} = \interp{D_2} \Leftrightarrow \interp{\lceil D_1 \rceil}= \interp{\lceil D_2 \rceil}$
		\end{center}
	\end{proposition}
	
	Notice also that $\mathrm{supp}(D) = \mathrm{supp}(\lceil D \rceil)$. This allows to reduce the study of general diagrams to diagrams of type $0\to n$ easier to handle. Thus, we only need to define a normal form for $0\to n$ diagrams.
	
	\begin{definition}[Normal form]
		A decohered ZX-diagram is in \textbf{normal form} if it is of the form: \input{nf.tikz}, where $A\in \mathcal{M}_{n \times k }(\mathbb{F}_{2})$ is an injective matrix in row reduced echelon form, $x\in \Im(A)^\perp $, $\lambda\in \mathbb{R}_{+}^{2^{k-1}}$ has full support, and $\Lambda \in \mathbb{R}_{+}^*$.
	\end{definition}
	
	The rules of the decohered ZX-calculus given in Figure \hyperref[fig3]{3} are powerful enough to put any diagram in normal form.
	
	\begin{proposition}\label{prop:diagtonf}
		Any decohered ZX-diagram $D:0\to n$ can be rewritten in normal form.
	\end{proposition}
	
	The interpretation of a normal form is: $\interp{\input{nf.tikz}}=\Lambda \sum_{y\in \mathbb{F}_{2}^k}\prod_{z\neq 0} \lambda_{z}^{z\cdot y} \ket{Ay+x} $. By Proposition \ref{prop:mapstate}, the normal form and Proposition \ref{prop:diagtonf} directly extend to diagrams of any type. So the interpretation of any decohered ZX-diagrams always has affine support. This provides a necessary criterion for a matrix to be representable by the decohered ZX-calculus. For example, the matrix representation of the AND gate is \scalebox{0.8}{$\begin{pmatrix}
			1 & 1 & 1 & 0 \\ 0 & 0 & 0 & 1
		\end{pmatrix}$} whose support is not affine, so we know that it cannot be represented by any decohered ZX-dagram. This condition is also sufficient. Indeed, the formula we gave for the interpretation of a normal form exactly parametrizes the set of vectors with affine supports.
	
	\begin{proposition}\label{prop:paramaff}
		For any vector $v\in \mathbb{R}_{+}^{2^n}$ with affine support, there is a unique injective matrix $A\in \mathcal{M}_{n \times k }(\mathbb{F}_2)$ in row reduced echelon form, a unique $x\in \Im(A)^\perp $, and a unique vector $u\in \mathbb{F}_{2}^k$ with full support such that: $v=\sum_{y} u_y \ket{Ay+x}$.
	\end{proposition}
	
	It follows that we can assign to each vector with affine support a unique decohered ZX-diagram in normal form.
	
	\begin{proposition}\label{prop:uniquenf}
		For any vector $v\in \mathbb{R}_{+}^{2^n}$ with affine support, there is a unique decohered ZX-diagram in normal form with interpretation $v$.
	\end{proposition}
	
	We now have everything needed to state our main final theorem.
	
	\begin{theorem}[Universality and Completeness]\label{thm}
		The decohered ZX-calculus is universal and complete for matrices in $\mathcal{M}_{2^m \times 2^n }(\mathbb{R}_+ ) $ with affine support.
	\end{theorem}
	
	Notice that a consequence of this result is that matrices with affine support form a prop $\textbf{AffMat}_{\mathbb{R}_+}$ corresponding exactly to the prop presented by the decohered ZX-calculus.
	
	\section{Research directions}
	
	As it is, the decohered ZX-calculus must be understood as another step in the understanding of the structure of discard ZX-calculus. The set of rules presented in \hyperref[fig3]{4} is still unsatisfactory, it would be interesting to replace Rule \hyperref[rule:L]{L} with a more efficient rule, similar to the role played by the Euler rule in the completeness of ZX-calculus \cite{vilmart2019near}. Also, the restriction to affinely supported maps, even if perfectly understandable from a theoretical point of view, is an important limitation in practice. Still, the decohered ZX-calculus provides interesting features opening new directions in the study of diagrammatic languages.
	
	\noindent\textbf{Hybrid classical/quantum reasoning:} The most obvious application of the decohered ZX-calculus is the representation of hybrid classical/quantum processes. Following \cite{borgna2021hybrid}, an important research direction is the development of languages able to depict classical control of quantum processes, it would allow for the extension of the work of \cite{borgna2022encoding} to more advanced quantum programming languages.
	
	\noindent\textbf{Normal form \textit{via} diagrammatic Fourier transform:} The current completeness proofs of the ZX-calculus are mostly unsatisfactory due to the \textit{ad hoc} nature of the normal forms, reproducing the coefficients of the matrices in a cumbersome way. On the contrary, our normal form naturally exploits the structure of ZX-calculus and the interactions between the computational basis and its Fourier transform. We expect more investigations on those diagrammatic transforms to allow for new normal forms and completeness proofs for diagrammatic languages.
	
	\noindent\textbf{Picturing random variables:} Finally, there is another way to interpret the decohered ZX-calculus that we did not explicitly develop in this paper. One can then see the decohered ZX-calculus as a compositional approach to affinely supported multivariate Bernoulli distributions studied in \cite{dai2013multivariate}. Applications of ZX-calculus-like languages to probability theory have still to be thoroughly investigated, in the line of \cite{stein2024graphical}. Probabilistic processes are very similar to the quantum one, they also require a matrix representation of exponential size and feature non-trivial correlations. One could expect that techniques that have been successful in the quantum case could also be relevant in the probabilistic setting.\newpage
	
	\sloppy
	\nocite{*}
	\bibliographystyle{eptcs}
	\bibliography{dec}
	
	\newpage
	\appendix
	
	\section{Proofs}\label{append}
	
	\textbf{Proof of Proposition \ref{prop:decpure}:}
	
	\begin{proof}
		For any $\rho$:
		
		\[\underline{A}^\circ \left(\rho\right) = \sum_{x,y} \dyad{y}A\dyad{x}\rho\dyad{x}A^\dagger \dyad{y} =\sum_{x,y} |A_{y,x}|^2 \dyad{y}{x}\rho\dyad{x}{y}= \uwave{|A|^2}\left(\rho\right).\]
	\end{proof}
	
	\textbf{Proof of Proposition \ref{prop:wavecomp}:}
	
	\begin{proof}
		
		The two first properties are straightforward:
		
		\[\uwave{I_n}= \underline{\sqrt{I_n}}^\circ = \underline{I_n}^\circ = \delta_n \qquad \quad \uwave{A}\otimes \uwave{B} = \underline{\sqrt{A}}^\circ \otimes \underline{\sqrt{B}}^\circ = \left(\underline{\sqrt{A}\otimes \sqrt{B}}\right)^\circ = \left(\underline{\sqrt{A\otimes B}}\right)^\circ = \uwave{A\otimes B}\]
		
		Composition is more subtle, for any $\rho$:
		
		\[(\uwave{A}\circ \uwave{B})(\rho) =\sum_{z,t} A_{z,t} \dyad{z}{t} \sum_{x,y} B_{y,x} \dyad{y}{x}\rho\dyad{x}{y}\dyad{t}{z}
		= \sum_{x,z} \sum_{y} A_{z,y}B_{y,x} \dyad{z}{x}\rho\dyad{x}{z}
		= \uwave{AB}(\rho).\]
	\end{proof}
	
	\textbf{Proof of Proposition \ref{prop:uniquefnf}:}

	\begin{proof}
		Lets assume that there is a non-zero $\Lambda\in \mathbb{R}_+ $ and a vector $\lambda \in \mathbb{R}_{+}^{2^n - 1}$ with full support such that:
		
		\[v=\interp{\input{fnf.tikz}}=\Lambda\sum_{x}\prod_{y\neq 0} \lambda_{y}^{x\cdot y} \ket{x}\]
		
		We directly have $v_0 = \Lambda $ and for each $z\neq 0$:
		
		\[\prod_{x} \left(\frac{v_{x}}{v_0}\right)^{\frac{-2(-1)^{x\cdot z}}{2^n}}=\prod_{x} \left(\frac{\Lambda}{v_0}\prod_{y\neq 0} \lambda_{y}^{y\cdot x}\right)^{\frac{-2(-1)^{x\cdot z}}{2^n}}=\prod_{y\neq 0} \lambda_{y}^{\frac{-2}{2^n}\sum_{x}(y\cdot x)(-1)^{x\cdot z}}=\prod_{y\neq 0} \lambda_{y}^{\delta_{y=z}} = \lambda_z \]
		
		Using $\sum_{x}(-1)^{x\cdot z}=2^n \delta_{z=0}$ and for any $y,z \in \mathbb{F}_{2}^{n}$:
		\[\frac{-2}{2^n}\sum_{x}(y\cdot x)(-1)^{x\cdot z}=\frac{-2}{2^n}\sum_{x}\left(\frac{1-(-1)^{y\cdot x}}{2}\right)(-1)^{x\cdot z}=\frac{1}{2^n}\sum_{x}(-1)^{x\cdot(y\oplus z)}-\frac{1}{2^n}\sum_{x}(-1)^{x\cdot z}=\delta_{y=z}-\delta_{z=0} \]
		
		So $\Lambda$ and $\lambda$ must be unique. To conclude, we check that:
		
		\[\Lambda\prod_{y\neq 0} \lambda_{y}^{y\cdot x}=v_0 \prod_{y\neq 0} \left(\prod_{z} \left(\frac{v_{z}}{v_0 }\right)^{\frac{-2(-1)^{z\cdot y}}{2^{n}}}\right)^{y\cdot x}=v_0 \prod_{z} \left(\frac{v_{z}}{v_0 }\right)^{\frac{-2}{2^{n}}\sum_{y}(x\cdot y)(-1)^{y\cdot z}}=v_0 \prod_{z} \left(\frac{v_{z}}{v_0 }\right)^{\delta_{z=x}-\delta_{z=0}}=v_{x}\]
	\end{proof}
	
	\textbf{Proof of Proposition \ref{prop:lfnf}:}
	
	\begin{proof}
		For the first diagram, using Rule \hyperref[rule:M]{M} we can turn the green spider into a red spider, then we use Rules \hyperref[rule:F2]{F2} and \hyperref[rule:M]{M} to fuse the red spiders and turn the resulting spider into green. Thus, we just have to show how to reduce green spiders to Fourier normal form. Constructing a vector $\mu$ of size $2^n -1 $ such that $\mu_y = \begin{cases}
			\lambda_i \text{ if } y=\{i\}\\ \text{ else } 1
		\end{cases}$ we get:
		
		\[\input{lfnf1.tikz}\]
		
		For the second diagram, we have:
		
		\[\input{lfnf2.tikz}\]
		
		For the third diagram, we have:
		
		\[\input{lfnf3.tikz}\]
		
		The case of the last diagram is more subtle:
		
		\[\input{lfnf4.tikz}\]
		
		Rule \hyperref[rule:L]{L} tells us that we can rewrite the upper part of the diagram in Fourier normal form: 
		
		\[\input{lfnf5.tikz}\]
		
		Now, identifying elements of $\mathbb{F}_{2}^{2^{2^n - 1} -1}$ with non empty sets of non empty sets and setting $\nu_x = \prod\limits_{\cup_{\ell\in s} \ell =x} \mu_s $ we get:
		
		\[\input{lfnf6.tikz}\]
		
		Finally:
		
		\[\input{lfnf7.tikz}\]
		
	\end{proof}
	
	\textbf{Proof of Proposition \ref{prop:perm}:}
	
	\begin{proof}
		It follows directly from the matrix equation $\mathfrak{s}_n A = \sigma_{A} \mathfrak{s}_n $.
	\end{proof}
	
	\textbf{Proof of Proposition \ref{prop:sanf}:}
	
	\begin{proof}
		We consider an affine diagram $D:0\to m $. We start by using Rule \hyperref[rule:V2]{V2} to ensure that all output wires are connected to decohered red spiders, then we apply Rules \hyperref[rule:F1]{F1} and \hyperref[rule:F2]{F2} to fuse all spiders having a neighbor of the same color. We end up with a bipartite graph that can be written as follows using scalable notations: 
		
		\[\input{sanf1.tikz}\]
		
		where $B\in \mathcal{M}_{n+m \times k}(\mathbb{F}_2 )$, $y\in \mathbb{F}^n$, $z\in \mathbb{F}^m$ and $t\in \mathbb{R}_+ $. Now setting $B=\begin{pmatrix}
			B_1 \\ B_2
		\end{pmatrix}$ we get:
		
		\[\input{sanf2.tikz}\]
		
		Decomposing $B_2 $ as $B_2 = P^{-1} \begin{pmatrix} I_r & 0 \\ 0 & 0
		\end{pmatrix}Q^{-1}$ where $r$ is the rank of $A_2 $ and $P$ and $Q$ are invertible: 
		
		\[\input{sanf3.tikz}\]
		
		Which is in the correct form.
	\end{proof}
	
	\textbf{Proof of Proposition \ref{prop:interpgaa}:}
	
	\begin{proof}
		Using Proposition \ref{prop:sanf}, we just have to compute $\interp{\input{sanf.tikz}}$ which gives the expected result.
	\end{proof}
	
	\textbf{Proof of Proposition \ref{prop:banf}:}
	
	\begin{proof}
		See Lemma 48 in \cite{booth2024complete}.
	\end{proof}
	
	\textbf{Proof of Proposition \ref{prop:diagtonf}:}
	
	\begin{proof}
		Given a decohered ZX-diagram $D:0\to n$ we start by using Rules \hyperref[rule:F1]{F1} and \hyperref[rule:F2]{F2} to extract all green and red parameters that are not $1$ or $0$. Then, using Rule \hyperref[rule:M]{M}, we turn the extracted red states into green states. This gives a diagram of the form:
		
		\[\input{dtonf1.tikz}\]
		
		where $L$ is an affine diagram. Using Proposition \ref{prop:banf} we can decompose $L$ and get a diagram of the form:
		
		\[\input{dtonf2.tikz}\]
		
		where $P$ and $Q$ are invertible matrices, $x$ and $y$ are binary words and $Q=\begin{pmatrix}
			Q_1 & Q_2
		\end{pmatrix}$. Using Propositions \ref{prop:perm} and \ref{prop:lfnf} we can rewrite the left part of the diagram in Fourier normal form and get a diagram of the form:
		
		\[\input{dtonf3.tikz}\]
		
		Finally, using Proposition \ref{prop:perm} we can act on $Q_2$ with invertible matrices and reduce it to its row reduced echeloned form $R$ using Gauss elimination. To complete our reduction we decompose $x=a+b$ with $b\in \Im(R)$ and $a\in \Im(R)^\perp $ and use Proposition \ref{prop:lfnf} to produce a preimage $c$ of $b$ from the left part of the diagram, we get: 
		
		\[\input{dtonf4.tikz}\]
		
		Which is in normal form.
	\end{proof}
	
	\textbf{Proof of Proposition \ref{prop:paramaff}:}
	
	\begin{proof}
		Given a vector $v\in \mathbb{R}_{+}^{2^n}$ with affine support, we know that $\mathrm{supp}(v)$ is an affine subspace of $\mathbb{F}_{2}^{n}$ so there it is of the form $\mathrm{supp}(v)= E + x $ where $E$ is a linear subspace of $\mathbb{F}_{2}^n $ and $x\in \mathbb{F}_{2}^n $. $E$ is unique and can be written as $E=\Im(A)$ for a unique injective matrix $A\in $ row reduced echelon form. To obtain $A$, take any matrix with image $E$ and apply Gaussian elimination. If we assume that there is an $x' \in \Im(A)^\perp $ such that $x+\Im(A) = x' + \Im(A)$ then $x+x' +\Im(A) = \Im(A)$, so $x+x' \in \Im(A)\cap \Im(A)^\perp = \{0\}$ and $x=x'$. Finally, any $z\in \mathrm{supp}(v)$ can be written as $z=Ay+x$ for a unique $y\in \mathbb{F}_{2}^k $, so setting $u_y = v_{Ay+x}$, we get:
		
		\[v=\sum_{z\in \mathrm{supp}(v)} v_z \ket{z}= \sum_y u_y \ket{Ay+x}\]
		
	\end{proof}
	
	\textbf{Proof of Proposition \ref{prop:uniquenf}:}
	
	\begin{proof}
		
		Given any vector $v\in \mathbb{R}_{+}^{2^n}$ with affine support, by Proposition \ref{prop:paramaff}, there is unique injective matrix $A\in \mathcal{M}_{n \times k }(\mathbb{F}_2)$ in row reduced echelon form, a unique $x\in \Im(A)^\perp $, and a unique vector $u\in \mathbb{F}_{2}^k$ with full support such that:
		\[v=\sum_{y} u_y \ket{Ay+x}\]
		
		Now, using Proposition \ref{prop:uniquefnf}, there is a unique non-zero $\Lambda \in \mathbb{R}_+$ and a unique vector $\lambda \in \mathbb{R}_{+}^{2^k - 1}$ with full support such that:
		
		\[u_y = \Lambda\prod_{z\neq 0}\lambda_{y}^{z\cdot y}\]
		
		Finally combining the Fourier normal form for $u$ with the affine diagram implementing the map $\ket{y}\mapsto \ket{Ay+x} $ we get the unique diagram in normal form with interpretation $v$.
	\end{proof}
	
	\textbf{Proof of Theorem \ref{thm}:}
	
	\begin{proof} $\quad$\\
		
		\noindent\textbf{Universality:} Given a matrix $M\in \mathcal{M}_{2^m \times 2^n }(\mathbb{R}_+ )$ with affine support. We consider the vector $m \in \mathbb{F}^{m+n}$ defined by $m_{x,y} = M_{x,y}$. By Proposition \ref{prop:uniquenf} there is a decohered ZX-diagram in normal for such that: 
		
		\[\interp{\input{nf.tikz}}=m\]
		
		Then:
		
		\[\interp{\input{univ.tikz}}=M\]
		
		So all matrices in $\mathcal{M}_{2^m \times 2^n }(\mathbb{R}_+ )$ with affine support can be represented by the discard ZX-calculus.\\
		
		\noindent\textbf{Completeness}: Given two diagrams $D_1 , D_2 :n\to m$ such that $\interp{D_1}=\interp{D_2}$, by Proposition \ref{prop:mapstate} this is equivalent to $\interp{\lceil D_1 \rceil}=\interp{\lceil D_2 \rceil}$. But by Proposition \ref{prop:diagtonf}, there are diagrams in normal form $N_1$ and $N_2$ such that $N_1 = \lceil D_1 \rceil $ and $N_2= \lceil D_2 \rceil$. Now, since $\interp{N_1}=\interp{N_2}$, and since $N_1$ and $N_2$ are in normal form, the uniqueness condition of Proposition \ref{prop:uniquenf} ensures that $N_1 =N_2 $. So $\lceil D_1 \rceil = \lceil D_2 \rceil$ and by Proposition \ref{prop:mapstate}, $D_1 =D_2$.
	\end{proof}
	
	\end{document}